\documentclass[journal,letterpaper]{IEEEtran_nssmic}

\usepackage{graphicx}  

\begin{document}

\title{The STACEE Ground-Based Gamma-Ray Detector}

\author{D. M. Gingrich,
L. M. Boone,
D. Bramel,
J. Carson,
C. E. Covault,
P. Fortin, 
D. S. Hanna, 
J. A. Hinton, \\
A. Jarvis,
J. Kildea, 
T. Lindner, 
C. Mueller,
R. Mukherjee, 
R. A. Ong,
K. Ragan, \\
R. A. Scalzo, 
C. G. Th{\'e}oret,
D. A. Williams, and 
J. A. Zweerink%
\thanks{Manuscript received November 15, 2004; revised April 21, 2005.
        This work was supported in part by the National Science
        Foundation, the Natural Sciences and Engineering Research
	Council, Fonds Qu{\'e}b{\'e}cois de la Recherche sur la Nature
	et les Technologies, the Research Corporation, and the
	California Space Institute.}%
\thanks{D.~M.~Gingrich is with the Centre for Subatomic Research,
        University of Alberta, Edmonton, AB T6G2N5, Canada and TRIUMF,
	Vancouver, BC V6T2A3, Canada (e-mail: gingrich@ualberta.ca).}%
\thanks{L.~M. Boone was with the Santa Cruz Institute for Particle
        Physics, University of California, Santa Cruz, CA 95064, USA.
        He is now with the Department of Physics, The College of
	Wooster, Wooster, OH 44691, USA.}%
\thanks{D.~Bramel and R.~Mukherjee are with the Department of Physics
        and Astronomy, Barnard College and Columbia University, New
	York, NY 10027, USA.}%
\thanks{J.~Carson, A.~Jarvis, R.~A.~Ong, and J.~A.~Zweerink are with the
        Department of Physics and Astronomy, University of California,
	Los Angeles, CA 90095, USA.}%
\thanks{C.~E.~Covault is with the Department of Physics, Case Western
        Reserve University, Cleveland, OH 44106, USA.}%
\thanks{P.~Fortin, D.~S.~Hanna, J.~Kildea, T.~Lindner. C.~Mueller, and
        K. Ragan are with the Department of Physics, McGill University,
        Montreal, QC H3A2T8, Canada.}%
\thanks{J.~A.~Hinton was with the Enrico Fermi Institute, University of
       Chicago, Chicago, IL 60637, USA.
       He is now with the Max-Planck-Institut f{\"ur} Kernphysik,
       D-69029 Heidelberg, Germany.}%
\thanks{R.~A.~Scalzo was with the Enrico Fermi Institute, University of
        Chicago, Chicago, IL 60637, USA. 
        He is now with the Lawrence Berkeley National Laboratory,
	Berkeley, CA 94720, USA.}%
\thanks{C.~G.~Th{\'e}oret was with the Department of Physics, McGill
        University, Montreal, QC H2A2T8, Canada.
        He is now with the Laboratoire de Physique, Coll{\'e}ge de
	France, F-75231 Paris CEDEX 05, France.}%
\thanks{D.~A.~Williams is with the Santa Cruz Institute for Particle
        Physics, University of California, Santa Cruz, CA 95064, USA.}%
}


\maketitle

\begin{abstract}
We describe the design and performance of the Solar Tower Atmospheric
Cherenkov Effect Experiment (STACEE) in its complete configuration.  
STACEE uses the heliostats of a solar energy research facility to
collect and focus the Cherenkov photons produced in gamma-ray induced
air showers.
The light is concentrated onto an array of photomultiplier tubes located
near the top of a tower.
The large Cherenkov photon collection area of STACEE results in a
gamma-ray energy threshold below that of previous ground-based
detectors. 
STACEE is being used to observe pulsars, supernova remnants, active
galactic nuclei, and gamma-ray bursts.    
\end{abstract}

\begin{keywords}
Gamma-ray astronomy detectors, Cherenkov detectors, Cosmic rays,
Triggering, Coincidence Detection, Delay circuit.
\end{keywords}

\section{Introduction}

\PARstart{G}{amma-ray} astronomy has become a very exciting area of research. 
Since 1991 the field has rapidly expanded due to the increased
quantity and quality of data as well as an improved theoretical
understanding of the related astrophysics.
The thrust in the field was primarily initiated by NASA's Compton
Gamma Ray Observatory (CGRO) and the ground-based detectors that ran
concurrently. 
The Energetic Gamma Ray Experiment Telescope (EGRET) aboard the CGRO
produced a catalog of over 200 high-energy point sources~\cite{Hartman}.
Since space-borne instruments are by necessity small detectors, they are
only able to detect sources below about 10~GeV due to flux limitations.
To increase the energy range, ground-based detectors must be used.

Almost all ground-based gamma-ray detectors use the atmospheric
Cherenkov technique. 
Typical Cherenkov telescopes detect gamma rays by using large steerable
mirrors to collect, focus, and image the Cherenkov light produced by the
relativistic electrons resulting from the interactions of high-energy
gamma rays in the upper atmosphere. 
This Cherenkov light is distributed on the ground in a circular pool
with a diameter of 200~m to 300~m.
The size of the light pool is almost independent of the primary
gamma-ray energy, but the Cherenkov photon density scales linearly with
the primary gamma-ray energy.
Imaging Cherenkov telescopes have a very large collection area relative
to satellite detectors, and need only capture a part of the total
Cherenkov light pool to detect the primary gamma ray.
This gives rise to a low-energy threshold of about 300~GeV.   

The energy range between EGRET and imaging Cherenkov telescopes remained
unexplored until recently because no detectors were sensitive to the
energy region between 10~GeV and 300~GeV.     
The wavefront sampling technique is a variant of the imaging Cherenkov
technique whereby the collecting mirror is synthesized by an array of
large steerable mirrors (heliostats) at a central-tower solar energy
installation. 
The large effective area of the collecting mirror allows one to trigger 
at lower photon densities, and therefore lower primary gamma-ray
energies. 
STACEE is a wavefront sampling detector designed to lower the threshold
of ground-based gamma-ray astronomy to approximately 50~GeV, near the
upper limit of satellite detectors~\cite{Hanna}. 
Three other projects of a similar nature have also been built:
CELESTE~\cite{Naurois}, Solar-2/CACTUS~\cite{Tripathi}, and
GRAAL~\cite{Arqueros}.   

STACEE is investigating established and putative gamma-ray sources.
One of its principle aims is to follow the spectra of active galactic
nuclei (AGN) out to energies beyond that of EGRET measurements to
determine where the spectra deviate from a power law.
Many EGRET sources are not detected by imaging telescopes despite the
fact that a simple extrapolation of EGRET spectra are often well within
the sensitivities of such detectors.
We must then conclude that the spectra are somehow attenuated in the
10~GeV to 300~GeV energy region.
This effect could be due to cut-off mechanisms intrinsic to the source, 
or to absorption effects between the source and the detector.
A likely absorption mechanism is pair production, wherein the
high-energy gamma ray combines with a low-energy photon (optical or
infrared) from the extragalactic background radiation field.
Absorption by pair production thus makes gamma-ray measurements of
distant sources an indirect method of measuring the integrated light
from past star formation. 

\section{STACEE Detector}

The STACEE detector was in a state of development from 1997 to 2002.
Stages of the construction were followed by observing periods using the
partially completed detector. 
Since spring of 2002, the detector has been fully completed and is being
used for astrophysical observations. 

\begin{figure}[htb]
\centering
\includegraphics[width=8cm]{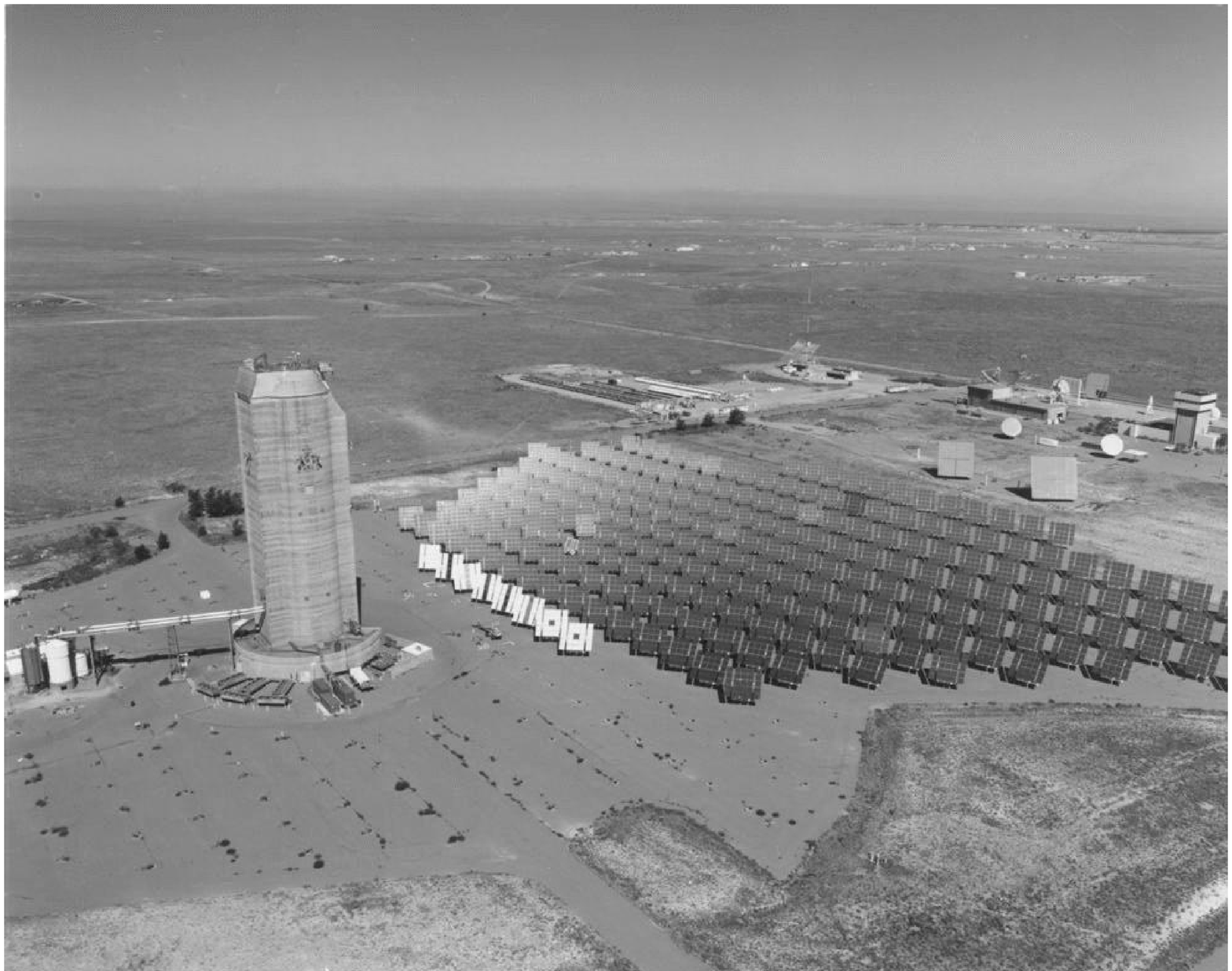}
\caption{National Solar Thermal Test Facility at Sandia National
Laboratories in Albuquerque, New Mexico.
The heliostat field covers an area of about 160~m by 260~m.}  
\label{nsttf}
\end{figure}

STACEE uses the National Solar Thermal Test Facility (NSTTF), which is
situated at Sandia National Laboratories in Albuquerque, New Mexico
(Fig.~\ref{nsttf}).  
Since the NSTTF is a solar research facility and STACEE observes at
night, there is no significant interference between the two programs.
Sixty-four of the 220 heliostats are used during clear moonless nights to
collect Cherenkov light from air showers and direct it onto five
secondary mirrors located near the top of a 61~m tower.
The secondary mirrors focus the Cherenkov light into groups of
photomultiplier tubes (PMTs) such that each PMT views a single
heliostat.  
Optical concentrators widen the aperture of each PMT and restrict its
field of view to reduce the number of night-sky background photons
detected.   
Signals from the PMTs are amplified and routed to a control room where
high-speed electronics measure the charge and relative arrival times of
the PMT pulses.
Signals above threshold are discriminated and processed by a delay and
coincidence trigger system.  

\subsection{Heliostats}

\begin{figure}[htb]
\centering
\includegraphics[width=\columnwidth]{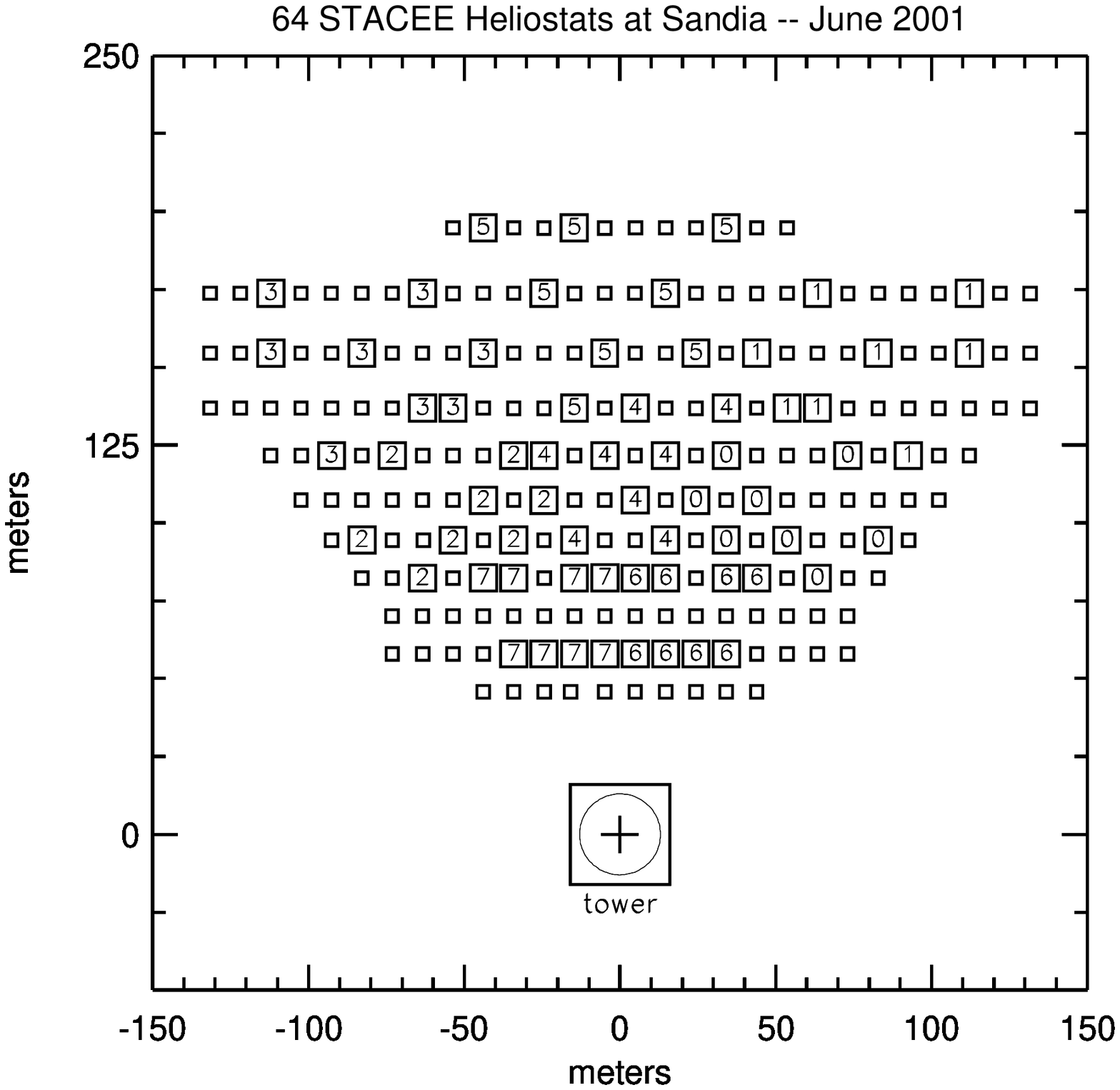}
\caption{Map of the heliostats in the National Solar Thermal Test
Facility field used by STACEE.
The STACEE heliostats are numbered according to the trigger cluster to
which they belong.
Clusters 0 and 1 correspond to the ``east'' camera, clusters 2 and 3
correspond to the ``west'' camera, clusters 4 and 5 correspond to the
``north'' camera, cluster 6 corresponds to the ``southeast'' camera, and
cluster 7 corresponds to the ``southwest'' camera.}
\label{map}
\end{figure}

Fig.~\ref{map} shows the heliostats in the NSTTF field that are used
by STACEE. 
The choice of the 64 heliostats used in STACEE is based on the desire
to uniformally sample the Cherenkov light pool expected from a shower 
impacting near the center of the array, while not crowding the PMTs in
the image plane. 
The collective area of the 64 heliostats is over 2300~m$^2$.

Each heliostat has a mirror area of 37~m$^2$, and consists of 25 square
facets mounted on a steel frame. 
Each facet is a 1.2~m by 1.2~m square of back-surfaced aluminized glass 
glued onto a thin metal sheet.
The facets are distorted into approximately parabolic shapes with the 
focus set to be equal to the distance to the tower. 
Each facet can be separately aligned so that their beams overlap at the
tower. 

The entire heliostat is mounted in a yoke structure which allows
rotation in azimuth and elevation angles.
The motion is achieved with two electric motors, each of which is
controlled by the NSTTF central computer using 13-bit encoders.

Facet alignment is checked and tuned using images of the Sun projected
onto the tower near solar noon (sunspots).
The Sun is a good diagnostic since its angular size ($0.5^\circ$) is
very similar to that of a Cherenkov shower.
The absolute pointing of each heliostat is calibrated to an accuracy of
0.05$^\circ$ using drift scans of bright stars. 

\subsection{Secondary Mirrors}

\begin{figure}[htb]
\centering
\includegraphics[width=8cm]{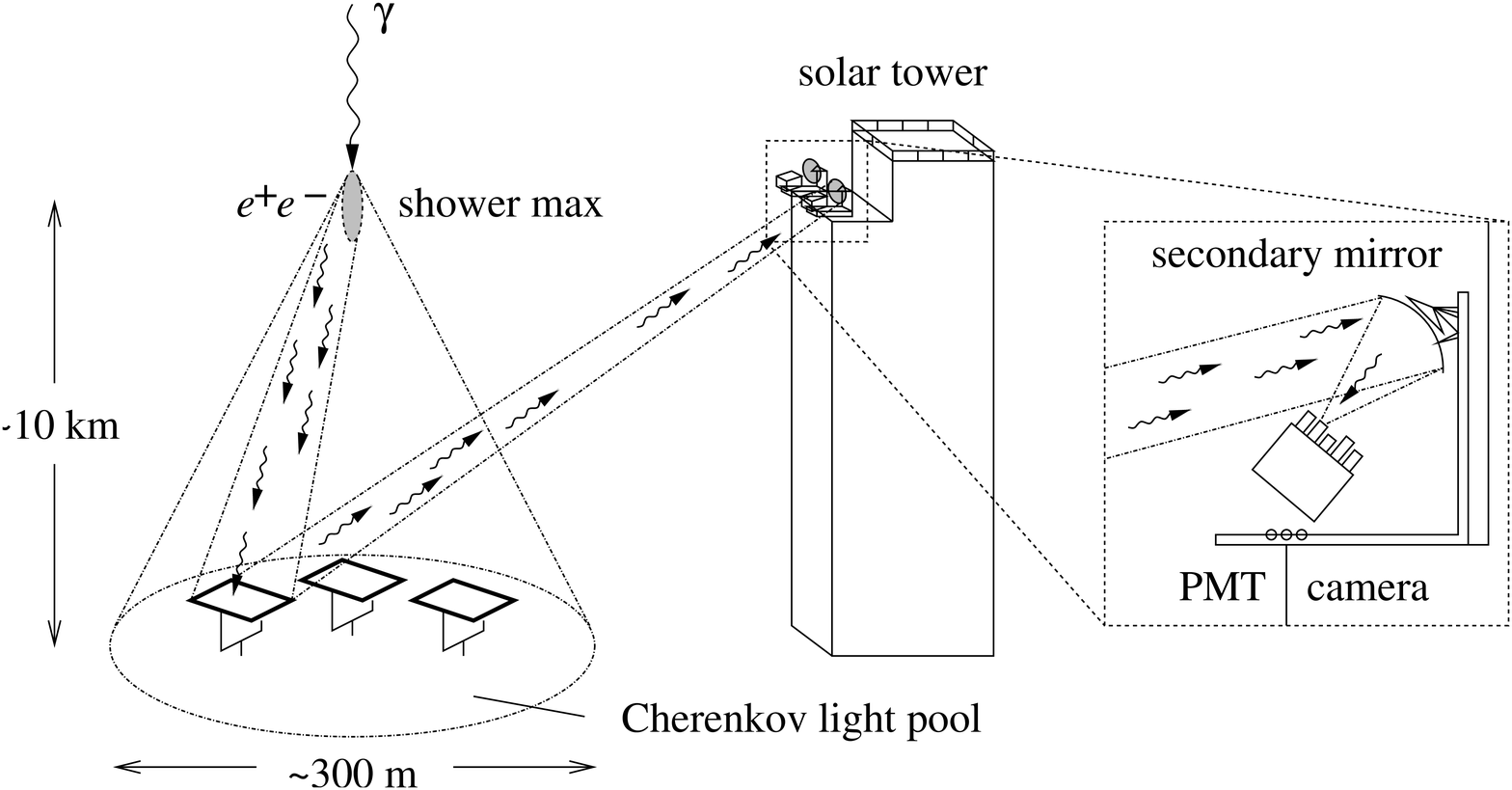}
\caption{Concept of the solar tower Cherenkov detection of gamma-ray air
showers (not to scale).}
\label{concept}
\end{figure}

Cherenkov photons are reflected by the heliostats onto five secondary
mirrors located near the top of the central tower (Fig.~\ref{concept}). 
Sixteen heliostats in the north, 16 heliostats in the east, and 16
heliostats in the west regions of the field (Fig.~\ref{map}) are
viewed by three independent mirrors located 49~m above the base of the
tower. 
Similarly, eight heliostats in the southeast and eight heliostats in
the southwest of the field (Fig.~\ref{map}) are viewed by two
independent mirrors 37~m above the base of the tower.

The three secondary mirrors at the 49~m level are spherical with a
nominal diameter of 1.9~m and a focal length of 2.0~m. 
Each is composed of seven identical hexagonal facets made from
front-surfaced aluminized glass in order to retain a high reflectivity
at ultraviolet wavelengths, where most of the Cherenkov light from air
showers is produced.
The two secondary mirrors at the 37~m level are single spherical mirrors
with a diameter of 1.1~m and a focal length of 1.1~m.

The secondary mirrors focus the light from the heliostats, which arrives 
as a wide beam, onto phototube assemblies fixed in position at the focal
plane. 
The optics are such that each heliostat is mapped onto a single PMT
channel. 
This one-to-one mapping is vital for pattern recognition, which is used
in trigger formation and background suppression.

\subsection{Cameras}

The final stage in the STACEE optics chain is the camera.
There is one camera for each secondary mirror.
The cameras at the 49~m level consist of 16 PMT assemblies and the
cameras at the 37~m level of eight PMT assemblies each.
Each PMT assembly consists of a PMT and light concentrator enclosed in a 
canister.
The PMT canisters are mounted in cylindrical sleeves attached to an
azimuthal-elevation mounting system secured to a slotted plate.
With this system, it is possible to position the PMT canisters anywhere  
laterally on the slotted plate and to adjust the orientation of the
canisters such that they point to the center of the secondary mirror.

The light concentrators are Dielectric Total Internal Reflection
Concentrators (DTIRCs)~\cite{DTIRC} made from solid UV-transparent
acrylic.  
These are non-imaging devices which use total internal reflection to
transport light from the front surface to the exit aperture.
The light from a circular area of 11~cm diameter is concentrated to an
exit diameter of less than 4~cm.
Only light from a given angular range can reach the exit aperture, so
the DTIRCs have the added feature of being able to define the field of
view of the PMT.

For far-away heliostats, spherical aberration distorts the shape of the
image and produces a long coma tail, large enough in some cases to
overlap the apertures of other DTIRCs.
While somewhat troublesome for certain calibration activities, this
overlap is not expected to present a difficulty during normal
astronomical observations; the arrival times of a Cherenkov wavefront at
the apertures of adjacent DTIRCs usually differ by several tens of
nanoseconds. 
Any crosstalk photons will therefore lie outside the coincidence trigger
window, and will not contribute to the trigger.  

\subsection{Photomultiplier Tubes}

STACEE uses the Photonis XP2282B photomultiplier tube with a
borosilicate window and a VD182K/C transistorized voltage divider.
This tube has a good sensitivity to short wavelengths (blue and UV),
where most of the Cherenkov light is concentrated. 
Each PMT views the light from a 37~m$^2$ heliostat so it generates
single photoelectrons from night-sky background at a rate in excess of
1.5~GHz.
The PMT rapid rise time of 1.5~ns and narrow output pulse width helps to
reduce pulse pile-up effects.
A small transit time spread of 0.5~ns results in a timing resolution of
the experiment of less than 1~ns. 
Excellent time resolution allows us to exploit the narrowness of the
Cherenkov wavefront at the trigger level to reject background from
showers produced by charged cosmic rays. 
Offline, good timing resolution is valuable in reconstructing the shape
of the wavefront (approximately spherical) in order to reject
background. 

The PMTs are supplied with high voltage from LeCroy 4032A high voltage
power supplies, which are controlled by a LeCroy  2132 CAMAC interface
in the control room. 
Voltages are typically in the neighbourhood of $-1600$~V.
The high voltage values are periodically adjusted to equalize the
response of all channels.

\subsection{Front-End Electronics}

\begin{figure}[htb]
\centering
\includegraphics[width=8cm]{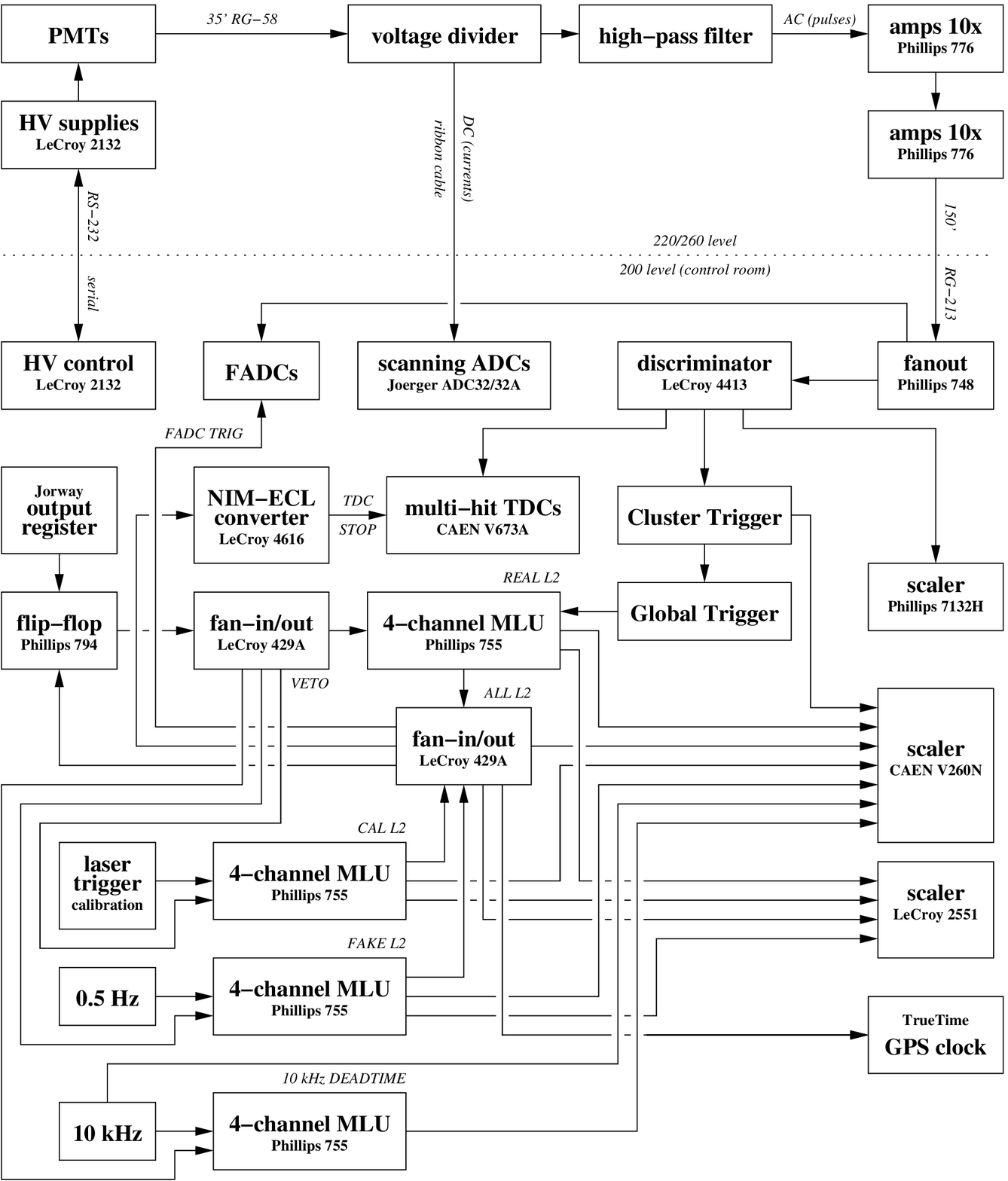}
\caption{Block diagram of the STACEE electronics.}
\label{electronics}
\end{figure}

Fig.~\ref{electronics} shows a block diagram of the STACEE electronics,
and Table~\ref{parm} summarizes the performance parameters that will be
discussed in the following sections.

\begin{table}[htb]
\caption{Performance parameters of the STACEE electronics and trigger
system.}
\centering
\begin{tabular}{|c|l|} \hline
Value & Description \\ \hline
0.5 ns      & PMT transit time \\
1.5 ns      & PMT rise time \\
2000 ns     & programmable delay range \\
1 ns        & precision on delay times \\
1-10 kHz    & cluster rates \\
5 Hz       & nominal global trigger rate \\
15.6 ns     & trigger signal jitter \\
8-12\%      & trigger deadtime \\
7.8-23.4 ns & cluster coincidence window \\
15.6 ns     & global trigger coincidence window \\
140 ns      & trigger latency \\
3.0 ns      & minimum pulse width that can be encoded \\
6 ns        & average double pulse resolution \\
15.6 ns     & mean width of double channel coincidence \\
1 GS/s      & FADC sampling rate \\
8-bit       & FADC resolution \\
1 V         & FADC dynamic range \\
1 ns        & experiment timing resolution \\
\hline
\end{tabular}
\label{parm}
\end{table}

Signals from the phototubes are filtered and amplified near the cameras
before being sent to the STACEE control room, located up to 18~m below
the detectors in the tower.
There they are discriminated, and used in timing measurements and trigger
logic. 
Concurrently, the analog pulses are continuously digitized.

The front-end analog electronics are physically close to the PMTs.
The PMT signals arrive at the front-end electronics via 11~m long RG58
cables. 
The signals are passed through high-pass RC filters having a time
constant of 75~ns. 
This filter blocks any DC component of the PMT signal and removes slow
PMT transients, which are not associated with Cherenkov signals. 

The pulsed component of the signals exiting the filters are amplified 
by two cascaded fixed-gain (x10) wide-band (275~MHz) amplifiers
(Phillips Scientific 776).
This amplification factor of 100 allows us to keep the PMT gain at
approximately $10^5$, which is expected to prolong the life of the PMTs
in an environment of high night-sky background light levels.

The filtered and amplified signals are routed through up to 40~m of
low-loss coaxial cables (RG213) from the detector levels to the control 
room level of the tower where they are fed into linear fanouts (Phillips
Scientific 748).  
The outputs of these fanouts are passed to the discriminators and flash
ADCs (FADCs). 

\begin{figure}[htb]
\centering
\includegraphics[width=\columnwidth]{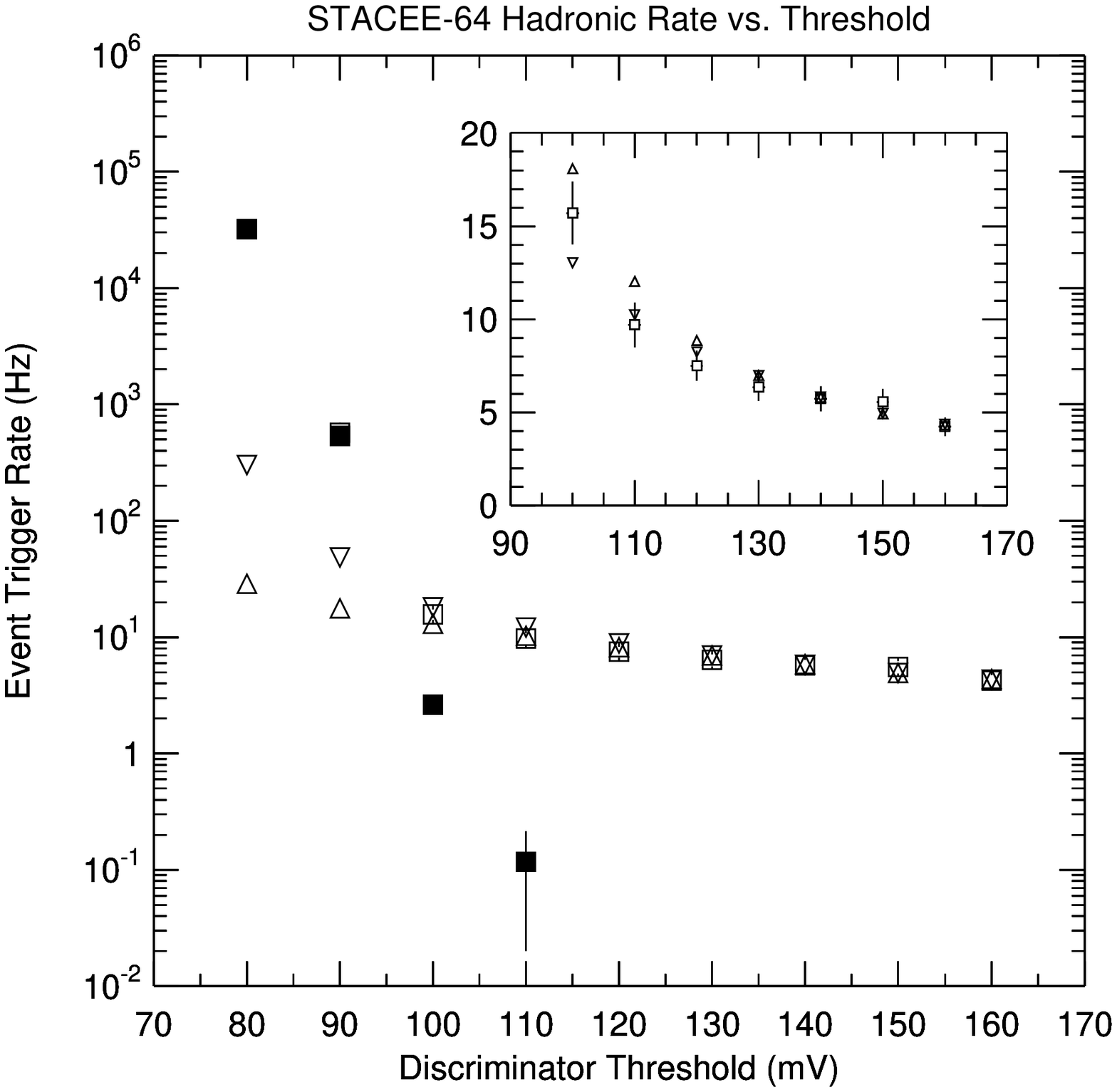}
\caption{Trigger rate versus discriminator threshold.
The empty squares are data taken with in-time delays and the filled
squares for random delays.
The triangles are for two simulation models: unit-slope (inverted) and
zero-intercept (upright) linear extrapolation. 
The insert shows the same points above a threshold of 90~mV, with a
linear vertical scale so that the error bars are visible.} 
\label{rvst}
\end{figure}

The analog signals from the PMTs are discriminated by 16 channel
discriminators (LeCroy 4413 or Philips Scientific 7106) operating with a
common threshold. 
The discriminator thresholds are set according to a rate versus
discriminator threshold curve like the one shown in Fig.~\ref{rvst}.
In this plot, one sees data for in-time delays (empty squares)
appropriate for Cherenkov triggers, and random delays (filled squares),
which show the contribution from accidental coincidences caused by
night-sky background photons. 
There is a breakpoint in the trigger rate, which is defined as the
discriminator threshold below which the rate climbs exponentially.
The location of the breakpoint depends on the individual channel rates,
the widths of the discriminator pulses, and the number of channels
required to form a trigger. 
At discriminator thresholds below the breakpoint, the rate is dominated
by accidental coincidences.
At very low discriminator threshold values the curve flattens due to
deadtime. 
At discriminator thresholds above the breakpoint, the rate decreases
slowly with discriminator threshold for in-time delays where the
experiment is triggering mostly on Cherenkov light.   
The operational discriminator thresholds were set 15~mV to 20~mV above
the breakpoint to minimize the background from accidental triggers. 
Background for STACEE consists almost entirely of hadron-initiated
Cherenkov events.  

The system should handle PMT discriminator rates up to about 10~MHz.
Typical discriminator rates are 1~MHz to 5~MHz.
The PMT discriminators contribute an effective deadtime that is rate
dependent, and is typically less than 5\%.

\subsection{Delay and Trigger System}

A unique challenge in STACEE, that led to the design of a custom-built
delay and trigger system~\cite{Martin}, is the requirement of dynamic
delays.  
Due to the Earth's rotation, the gamma-ray source appears to move
across the sky during the course of a night's observations.  
This effect continously changes the relative arrival times of the
Cherenkov photons at each heliostat.
In order to maintain tight timing coincidences, signals from different
channels are delayed by different amounts to correct for the source
movement. 

The required range of delay times is determined by the geometry of the
heliostat array and the maximum zenith angle at which we observe a
source. 
The programmable delay system has sufficient range (approximately
2~$\mu$s) to trigger on Cherenkov showers coming from any region of the
sky within a cone of 90$^\circ$ opening angle around the zenith. 
Individual delay settings can be controlled with a nominal precision of
1~ns.
To ensure precise timing, every channel is calibrated with test pulses.
The delays are updated every few seconds as STACEE tracks a source
across the sky.
Typical changes in the delays are approximately 1~ns per 15~s of elapsed
time. 

STACEE has a two-level trigger system.
The 64 heliostats are divided into eight clusters of eight heliostats
each (Fig.~\ref{map}).  
The discriminator outputs from the eight channels in each cluster are
routed through delays programmed to bring in-time hits into
coincidence. 

The number of coincident PMTs in a cluster, and the number of coincident
clusters, are chosen to optimize the quality factor for the rejection of
hadronic air showers according to Monte Carlo simulations. 
The discriminator threshold is then set at a level which makes the
overall event trigger rate from chance coincidences due to fluctuations
in the night-sky background photons negligible (less than
0.2~min$^{-1}$). 

Typical trigger settings are a discriminator threshold of 140~mV (about
5.5 photoelectrons), five out of eight PMTs are required to form a
cluster trigger, and five out of eight cluster triggers are required to
form a global trigger.  
At these trigger settings Cherenkov events are recorded at a rate of
about 5~Hz for offline analysis.
Typical cluster trigger rates are 1~kHz to 10~kHz. 
The precise width of the cluster coincidence window as applied to a
given series of pulses varies between 7.8~ns and 23.4~ns due to the
implementation of the trigger logic~\cite{Martin}; the mean width for a
double-channel coincidence is 15.6~ns. 
The coincidence window for the global trigger is 15.6~ns.
The rate of false triggers, due to spurious coincidences of night-sky
background light, decreases dramatically as the coincidence window is
made shorter. 
The physical limit of the coincidence window is given by the intrinsic
arrival-time width of the Cherenkov wavefront of about 4~ns.

A by-product of the two level trigger system is that the light pool is
required to be spread out over the entire heliostat array.
This feature is typical for showers induced by gamma rays and is used to
distinguish them from showers initiated by charged cosmic rays.

The rate of accidental triggers due to random night-sky hits in the PMTs
can be directly determined by using random settings for the programmable 
delays (Fig.~\ref{rvst}), which in effect is like pointing the
heliostats in random directions. 

The global coincidence triggers are combined with free running trigger
signals generated at a rate of 0.5~Hz. 
Free running triggers are used for determining individual channel rates
and pedestal values.
Whenever either type of trigger occurs (coincidence or free running), a
common stop signal is sent to latch a GPS clock time and initiate the
event read out. 
During readout, a veto is asserted to prevent the occurrence of additional
triggers.
The veto is cleared by the data acquisition (DAQ) program at the
conclusion of readout.

\subsection{Implementation of the Delay and Trigger System}

The STACEE delay and trigger system digital logic circuit is based on
Altera FLEX 10KE embedded programmable logic devices.
The system consists of 10 boards in VME format: one clock distribution
board, eight 8-channel cluster trigger boards, and one global
coincidence trigger board.  
The system makes extensive use of TTL signals.
The only ECL components are ECL/TTL translators, used to receive the
input signals from the discriminators, and to receive and transmit
signals between the cluster trigger boards and the global trigger board.

\subsubsection{Clock Board}

The delay and trigger system is a synchronous digital logic circuit
except for the tapped delay lines at the inputs. 
The system clock is generated by a separate clock board containing a
32~MHz TTL clock oscillator.
The clock is distributed over the SERCLK line on the VME bus to the nine 
other boards in the system.
Each of these boards uses the SERCLK to synchronize two local
phase-locked loop chips to generate 128~MHz and 64~MHz clocks.   

\subsubsection{VME Interface}

All communication with the system is over the VME bus.
The boards are started and stopped by writing to a control register on
each board.
The programmable delay values are written to the boards by first
selecting the channel to receive the data and then writing to that
channel's delay register. 
The delay specified for each channel can be changed in real time with the
system running, as long as the change in the delay value is less than
15.6~ns from the previous value. 
The programmable delay register can not be read.
The trigger conditions are set by writing the multiplicity requirement
and coincidence mode to a register on each board.
After a global coincidence trigger is issued, the timing information for
each channel for the 31.25~ns period preceeding the trigger can be read.
In addition, control, status, and test registers are readable and
writable.

\subsubsection{Time Encoding Logic}

The differential ECL signals from the discriminators are converted to
single-ended TTL signals on each cluster trigger board. 
An Altera MAX 700 complex logic device (EPM7032) 
is used to select eight inputs corresponding to a unique set of
heliostats from the 16 inputs on the board.  

The 128~MHz clock frequency allows for only a resolution of 7.8~ns in
the delays.
To obtain approximately 1~ns resolution in the delays, tapped delay
lines are used in front of the purely digital circuit. 
Monolithic 10-tap fixed delay lines (3D7110) from Data Delay Devices
are used. 
The tap-to-tap delay of these lines is $(1.0\pm 0.5)$~ns.
The input to the first tap delay will lead to an average 1~ns common
delay on all channels which just adds to the trigger latency.
The constant channel-to-channel variation of this initial delay can be
measured and corrected for by including them in  the relative
programmable delay values.  

The time encoding logic runs in a EPF10K30E embedded programmable logic 
device at a clock frequency of 128~MHz. 
The lower eight tap outputs are latched every 7.8~ns.
As the resulting bit pattern is not synchronous with the 128~MHz clock,
a four stage pipeline follows the input flip-flop to recover from
metastability.  
By examining the pattern of bits it is possible to determine the rising
edge of the TTL input pulse relative to the previous clock edge. 
Each 7.8~ns, when the taps are latched, the time can be coded into 3
bits which we call a vernier code.  
To determine the rising edge of the input pulse, five consecutive tap
values are examined for a pattern -- two or more zeros followed by at
least three ones -- that is consistent with the minimum pulse widths and  
double-pulse resolutions of the 3D7110 tapped delay lines, and the LeCroy
4413 or Philips Scientific 7106 discriminators. 
Since the input pulses are not synchronized with the clock, the 5-bit
pattern can occur across a clock edge so the tap outputs from three
consecutive clock periods must be examined to unambiguously determine the
hit pattern and its preceeding clock edge.  
If a valid hit pattern is detected, a fourth bit is set and added to the
3-bit vernier code.
The time encoding has a minimum latency of approximately 47~ns.
The minimum pulse width that can be encoded is approximately 3.0~ns.
The average double-pulse resolution is approximately 6~ns.
Pulses separated by 8~ns will be resolved with full efficiency.

\subsubsection{Delay Logic}

Since the asynchronous arrival times of the input pulses are time
encoded, the input signals are not delayed but rather the vernier codes
are delayed in first-in first-out (FIFO) memories,
Since the vernier codes span a range of 8~ns, the FIFO memories should
be clocked at 125~MHz.
As this was rather fast for field programmable gate arrays available at
the time, we choose to multiplex the vernier codes into an EPF10K100
embedded programmable logic device clocked at 64~MHz. 
In order to synchronize the two programmable logic devices clocked at
different frequencies, the 4-bit vernier codes produced by the EPF10K30
are split into two separate paths on alternate 128~MHz clock rising
edges, and then resynchronized at 64~MHz, as 8-bit vernier code pairs.

The value of a 7-bit reference counter, in the clock cycle in which the
vernier code pair would be pushed onto the FIFO memory, is appended to
the vernier code bits.
This produces for each 15.6~ns clock period effectively two 11-bit time
stamps sharing the same reference counter value, including two code
validation bits corresponding to the 0 to 7~ns range and the 8~ns to
15~ns range of the input pulse arrival times.

The time stamp must now be delayed in the FIFO memory for a specified 
time given by the value in a programmable delay register.
For each channel, the content of its programmable 11-bit delay register
is combined with the time stamp to produce the delay-time stamp.  
The delay-time stamp is pushed onto a FIFO memory embedded in the
EPF10K100.  
There is one FIFO memory for each of the eight channels.
The upper seven bits of the delay-time stamp of the next available
output data in the FIFO memory is compared continously with the
reference counter. 
When there is a match, the vernier codes of the current delay-time 
stamp are popped from the FIFO memory and inserted into a coincidence
pipeline. 
In this way, each hit for a channel is kept in the corresponding FIFO
memory for a period equal to the programmed delay register.
The FIFO memory is 2000~ns deep.

Each time the vernier codes are inserted into the coincidence pipeline
the vernier codes from the next delay-time stamp are also inserted into the
coincidence pipeline.
In this way there is 31.25~ns of vernier code information available in
the pipeline from which to form the cluster trigger decision. 

\subsubsection{Cluster Trigger Logic}

The cluster trigger logic inspects the vernier codes of the eight channels
and looks for a coincidence within a selectable time window.
The cluster trigger coincidence logic has two different modes of
operation: a wide-coincidence mode and a narrow-coincidence mode.
In wide mode, a hit anywhere in the first 23.4~ns interval starting
at the edge of the 64~MHz clock -- that is, within the first three 4-bit
vernier codes, each covering one 7.8~ns interval -- is considered to be
in the coincidence window. 
A cluster trigger is produced if the number of channels which have such
a hit is greater than or equal to the required cluster multiplicity
value stored in the multiplicity register on the cluster trigger board.

In narrow-coincidence mode, a coincidence window slides through the
vernier codes in 1~ns increments, starting at the 64~MHz clock edge, and
a cluster trigger is produced if the number of channels with at least
one hit in this window exceeds the multiplicity requirement set in the
multiplicity register on the cluster trigger board.

The width of the coincidence window for narrow-coincidence mode is fixed.
It can be changed by modifying the EPF10K100 circuit and reloading the
configuration file. 
At present the narrow-coincidence mode window width is 12~ns.
If the coincidence window is larger than 7.8~ns, a channel can have more
than one hit within the coincidence window.

The cluster trigger signal is synchronized with the 64~MHz clock and fed
to the global trigger board for the duration of one clock period.
The cluster boards also provide NIM pulses on the front panels, which are
counted by a VME scaler (Caen V260N).

Because of the latency associated with transferring the cluster trigger
signals to the global trigger boards, the cluster trigger boards do not
stop if a local cluster trigger is asserted.
Instead, they keep the vernier hits in a pipeline and only stop upon
reception of a ``hold'' signal from the global coincidence trigger board.
If the hold signal is received, the vernier hits for the last 31.25~ns
are extracted from the coincidence pipelines and made available for
readout.
The overall trigger latency is approximately 140~ns.

Notice that the cluster trigger logic is performed every 15.6~ns,
although it uses up to 31.25~ns of vernier code information.
Thus there is some redundancy, and a single ``event'' can cause
consecutive clock cycles to generate cluster triggers.
As a result, the cluster trigger output signal is sometimes 15.6~ns wide
and sometimes 31.25~ns wide.

\subsubsection{Global Trigger Board}

The cluster trigger information is routed to the global trigger board
over front-panel connectors using differential ECL signals.
The global trigger logic simply forms the arithmetic sum of the number 
of cluster triggers it receives within one 15.6~ns clock period.
If this sum is equal to or larger than the value in the global trigger
multiplicity register, then a global coincidence trigger is asserted.

The assertion of the global trigger generates a hold signal that is
fanned out to the cluster boards over front-panel connectors using
differential ECL signals.
In addition, a NIM-level trigger signal from the global trigger board
alerts the DAQ that the event should be read out. 
At this stage, four 4-bit vernier codes, corresponding to the 31.25~ns
period preceeding the global trigger, are extracted from the coincidence
pipeline, and read over the VME bus for every board and for every
channel.
Once all the required information has been gathered, the DAQ software
issues a clear-FIFO command to all the cluster trigger boards, and
reactivates the trigger by writing into the control register of the
global coincidence trigger board. 

\subsubsection{Performance}

Since the input pulses arrive asynchronously with respect to the clock,
even after the programmable delays are applied, the trigger signal will
have a 15.6~ns jitter.
The jitter can be corrected for offline by using the vernier time codes
of the individual hits.

The coincidence windows are discreet with a 1~ns step.
Thus, the coincidence efficiency as a function of the difference in time
between two channels is a trapezoid with a base at zero efficiency that
is 2~ns longer than the top at 100\% efficiency.
There are departures from the ideal case due to the fact that the tapped
delay lines used for the vernier encoding are not perfect.
The broadening of the coincidence resolution relative to the ideal
value is significantly less than 1~ns.
This includes contributions from all sources of broadening: unequal
delay taps, encoding eight 1~ns tap delays in a 7.8~ns clock period,
tolerance in the absolute values of the delays, etc. 

\subsection{Flash ADC System}

To reconstruct the energy and direction of the primary gamma ray, a
commercial FADC system made by Acqiris Inc. is used.
Sixteen channels of FADCs are contained in a special crate along with
their own embedded computer running a version of the Linux operating
system, modified to support real-time applications.
Four FADC crates make up the system.

Each electronics channel is sampled at 1~GS/s with an 8-bit resolution
and dynamic range of 1~V. 
The zero points of the FADC inputs are calibrated to a precision of 1~mV
RMS, and the channel-to-channel gains of the system are equalized to
within 0.5\%.


The fully digitized PMT waveforms allow not only accurate measurements
of the timing and intensity of the wavefront, but also the measurement
of the charge-timing correlations, such as the distribution of Cherenkov 
photon arrival times at each heliostat. 
This enables us to use various new methods~\cite{Scalzo,Zweerink} to
reject the large background of hadronic events due to charged cosmic
rays, while  retaining gamma-ray initiated events.  
The FADCs are also routinely used to calibrate and monitor the gains of
the PMTs using a custom-designed laser calibration system.

\subsection{Laser Calibration System}

STACEE is equipped with a laser calibration system comprising a
100~$\mu$J nitrogen laser and dye cell feeding a network of optical
fibres through a system of adjustable neutral density
filters~\cite{laser}. 
The fibres deliver light to the PMTs by exciting small diffuser plates
attached to the center of the secondary mirrors. 
The intensity of each laser shot is measured independently using four PIN
photodiodes. 

The system is used for measuring the relative time differences between
PMT channels.
It is also used for monitoring gain stability of the PMTs.

The timing resolution is estimated by examining distributions of
residuals to shower fits.
It is better than 1~ns for all channels and does not depend on the
pointing angle.
It is stable over time.

\subsection{Miscellaneous Electronics}

STACEE uses a GPS clock to provide a time-stamp, accurate to 1~$\mu$s,
for all recorded events.
These time-stamps are necessary for pulsed emission searches.

Counters are used to measure the deadtime due to the readout.
Whenever a trigger occurs, a flip-flop is set which asserted a veto
signal that inhibits further triggers.
This veto is required to prevent additional triggers from interfering
with the readout, and it is cleared by the DAQ computer once the readout
is complete. 
A 10~kHz clock is sent to two scaler channels to measure the deadtime.
One scaler counts a vetoed copy of the clock and the other counts an
unvetoed copy.
The ratio of the vetoed scaler count to the unvetoed scaler count gives 
the livetime fraction of the experiment.
Scalers are also used to monitor the rates of all the discriminated PMT 
pulses as well as all the cluster triggers. 

The cluster trigger deadtime during normal running is negligible.  
The overall deadtime is dominated by the readout time and varied
between 8\% and 12\% depending on the trigger rate.

\subsection{Data Acquisition}

The data acquisition system for STACEE is based on a VME embedded PC which
reads out the VME crate and a CAMAC crate via a branch bus.
Data are read out after each trigger and stored on a local disk.
At intervals, the data are copied from the local disk, along with the
FADC data, to a separate PC and eventually written to DLT tape for
archiving and off-site analysis. 
Monitoring in real time occurs on a PC running Linux.

In the NSTTF control room, the positions and motions of the heliostats
are monitored and the data are written to disk to be merged with the
detector data at the end of observation.
In a similar fashion, the weather and atmospheric conditions are
monitored and recorded.


\section{Detector Modeling}

The modeling of STACEE consists of simulating extensive air showers,
the optical throughput of the detector, and the electronics.    
Since STACEE is an atmospheric Cherenkov experiment, the atmosphere is
an integral part of the detector.

\subsection{Extensive Air-Shower Simulations}

The design and understanding of STACEE is aided by the CORSIKA
air-shower simulation package~\cite{CORSIKA}. 
CORSIKA makes use of packages such as EGS4~\cite{EGS} and
GHEISHA~\cite{GHEISHA}, which are widely used in other scientific 
fields.
CORSIKA simulates the entire development of an extensive air shower,
starting with the first interaction of the primary particle in the upper
atmosphere, and follows all generated secondary particles until they
reach the ground or their energy falls below the point where they no
longer contribute to shower development.
For the energies relevant to STACEE ($E < 10^{12}$~eV) we are able to
follow all particles since the multiplicities are small enough. 

The intervening processes accounted for by the program include
ionization, bremsstrahlung, and pair production, as well as effects such
as the deflection by the Earth's geomagnetic field, and Coulomb
scattering in the atmosphere. 

The development of the shower depends on the density profile of the
atmosphere, and thus the assumed atmospheric profile is an important
input of the simulation.
The rate and angle of production of Cherenkov photons by particles in
the shower depend on the local refractive index.
In addition, the attenuation of Cherenkov light due to Rayleigh and Mie
scattering and absorption by oxygen allotropes are important simulated
effects. 

\subsection{Optical Simulations}

The second part of the STACEE simulation chain traces the optical path
of Cherenkov photons through the detector optical elements.
For this part, a custom-written ray-tracing package called ``sandfield''
(Sandia Field Simulator) was developed.
Sandfield follows the path of every Cherenkov photon through the
optical elements (heliostats, secondary mirrors, and DTIRCs) onto the
PMT photocathodes, folding in transfer efficiencies at every stage.
The arrival times are smeared with a Gaussian resolution of 0.5~ns width
equivalent to the transit time spread of the PMT.
The end result is a list of photoelectrons and their arrival times at
the PMTs per channel.
These lists are passed to the electronics simulator for further
processing. 

\subsection{Electronics Simulations}

STACEE PMTs are bombarded with a high flux of photons either from air
showers or from night-sky background light.
The elevated rates caused by pile-up effects need to be understood
quantitatively, so a detailed electronics simulation package is
essential. 

The photoelectrons from the air shower (generated by the sandfield
program) are combined with random night-sky background photoelectrons
generated uniformly in time according to Poisson statistics.
A simulated analog waveform is built up by superimposing single
photoelectron pulses using an analytical pulse shape and adding them
according to their arrival times. 

The different gains of the channels are simulated by appropriate scaling of
the pulse amplitudes. 
The waveforms are finally passed through a simulation of the
discriminators. 

\subsection{Simulation Cross-Checks}

In order to verify the simulations, calibration data are recorded and
compared with the simulations.
CORSIKA was compared to another commonly used extensive air-shower
simulation package, MOCCA~\cite{MOCCA}.
The simulations agreed at the 20\% level.
We have chosen CORSIKA because it simulates the air-shower processes
in greater detail, particularly in the simulation of hadronic primaries.

For sandfield, cross-checks have included simulating the sunspot data
and the drift-scans of stars.
For the electronics simulation, most quantities read out in the real
data have been cross-checked, including PMT rates, distributions of
FADC trace samples, pulse-height distributions, and the overall trigger
rates, as functions of photocurrents, gains, discriminator thresholds,
etc.   

A powerful check of the simulation is to reproduce the absolute rate
versus threshold curves for various PMT channels.
The PMT rate at low threshold is due to the random pile-up of single
photoelectrons exceeding the discriminator threshold.
It is directly related to the photocurrents, both quantities depending
on the single photoelectron rate, the PMT gain, and the average pulse
width.
Small variations in any of these quantities produces large changes in
the rate.
Thus, reproducing actual rate versus threshold curves is a sensitive
test of the validity of the simulation, particularly the calibration of
the discriminator threshold level in terms of photoelectron
equivalents.
This calibration directly affects the gamma-ray energy threshold
calculation. 

Fig.~\ref{rvst} shows a comparison of these simulations with a real-data
rate versus threshold curve. 
The simulations are clearly inaccurate below a threshold of 100~mV.
However, this is approximately the point at which accidental
coincidences begin to dominate the event trigger rate, as is visually
apparent by the sharp break in slope.
Since the simulation does not currently support a mode which can trigger
on night-sky background as well as simulated Cherenkov showers at the
same time, it is unrealistic to expect it to simulate these points
correctly. 
However, at thresholds of 100~mV and above, the simulations perform
quite well.

A good end-to-end test of the simulation chain is the calculation of the
rate due to cosmic-ray triggers.
The cosmic-ray spectrum in the energy range relevant to STACEE is well
known.
Cosmic-ray data are obtained by collecting showers from the zenith.
Many cosmic-ray runs are taken by STACEE so a comparison between
calculated and measured cosmic-ray rates is straight-forward.
A comparison of real versus simulated hadronic cosmic-ray event rate
for several detector headings (hour angles) agree to 5\% or
better~\cite{Scalzo2}. 

Although past published results from STACEE include only integral
fluxes, more information can be extracted from the measured charges and
arrival times of the PMT pulses associated with the Cherenkov wavefront.
These event reconstruction techniques have been tested only in
simulations.
Although much work remains to be done, the most realistic simulations
performed to date show great promise in the new reconstruction
techniques~\cite{Scalzo2}.
The shower-core resolution is 10~m, independent of energy for all
showers with energies between 300~GeV and 2~TeV.
The energy resolution is about 20\% if the core is known to within 10~m
of its actual location.

\section{Summary}

STACEE is a complete ground-based Cherenkov wavefront sampling gamma-ray
telescope using heliostat mirrors of a solar energy research facility. 
Cherenkov light from air showers generated by the impact of high-energy
gamma rays on the upper atmosphere is collected with a set of 64
heliostats with a collective area of over 2300~m$^2$.
To date STACEE has achieved a low-energy threshold of about 100~GeV, 
which is lower than previously obtained by ground-based imaging
detectors. 
Since the complete detector came into operation in spring of 2002,
the measuring time was spent for observations of the Crab nebula, the
active galactic nuclei 3C66A, OJ+287, W-Comae, Markarian 421 and H1426,
and five gamma-ray bursts.

A modification to the STACEE electronics system during the summer of
2004 to move the FADC system closer to the PMTs and to install new
high-gain pre-amplifiers at the PMTs will allow STACEE to operate with a
faster, cleaner, electronic system at a lower energy threshold during
the upcoming seasons.

\section*{Acknowledgment}

We are grateful to the staff at the National Solar Thermal Test Facility
for their excellent support.





\begin{thebibliography}{99}

\bibitem{Hartman}
R.~C.~Hartman, D.~L.~Bertsch, S.~D.~Bloom, A.~W.~Chen, P.~Deines-Jones,
and J.~A.~Esposito \textit{et al.}, ``The third EGRET catalog of
high-energy gamma-ray sources," \textit{Astro. J. Suppl.}, vol. 123,
pp. 79-202, July 1999.

\bibitem{Hanna}
D.~S.~Hanna, D.~Bhattacharya, L.~M.~Boone, M.~C.~Chantell, Z.~Conner,
and C.~E.~Covault \textit{et al.}, ``The STACEE-32 ground based gamma-ray
detector,'' \textit{Nucl. Instrum. Meth.} vol. A491, pp. 126-151, Sept. 2002. 

\bibitem{Naurois}
M.~de~Naurois and CELESTE Collaboration, ``Measurement of the Crab
flux above 60~GeV with the CELESTE cerenkov telescope,''
\textit{Astro. J.}, vol. 566, pp. 343-357, Feb. 2002. 

\bibitem{Tripathi}
S.~M.~Tripathi, D.~Bhattacharya, J.~Lizarazo, P.~Marleau, G.~Mohanty,
and U.~Mohideen \textit{et al.}, ``The Keck Solar Two gamma-ray telescope
and its observations of the Crab Nebula,'' in \textit{Proc. ASS 200th
meeting}, 25.03, Albuquerque, NM, June 2002, p. 676.

\bibitem{Arqueros}
F.~Arqueros and GRAAL Collaboration, ``The GRAAL experiment,"
\textit{Nucl. Phys. B--Proc Suppl.}, vol. 114, pp. 253-257, Feb. 2003.

\bibitem{DTIRC}
X.~Ning, R.~Winston, and J.~O'Gallagher, ``Dielectric total internal
reflecting concentrators,'' \textit{Applied Optics}, vol. 26,
pp. 300-305, Jan. 1987. 

\bibitem{Martin}
J.-P.~Martin and K.~Ragan, ``A programmable nanosecond digital delay and
trigger system,'' in \textit{IEEE NSS/MIC Conf.}, Lyon, France, Vol. 2,
Oct. 2000, pp. 12/141-12/144.

\bibitem{Scalzo}
R.~A.~Scalzo and STACEE Collaboration, ``Optimized pointing strategies
for solar tower ACTS,'' in \textit{Proc. 28th International Cosmic
Ray Conference}, OG 2.5, Tsukuba, Japan, July 2003, pp. 2799-2802.

\bibitem{Zweerink}
J.~Zweerink and STACEE Collaboration, ``Using GHz FADCs to reject
hadrons from STACEE data,'' in \textit{Proc. 28th International Cosmic 
Ray Conference}, OG 2.5, Tsukuba, Japan, July 2003, pp. 2795-2798.

\bibitem{laser}
D.~Hanna and R.~Mukherjee, ``The laser calibration system for the STACEE
ground-based gamma ray detector," \textit{Nucl. Instrum. Meth.} vol. A482,
pp. 271-280, Apr. 2002.

\bibitem{CORSIKA}
D.~Heck, J.~Knapp, J.~N.~Capdevielle, G.~Schatz, and T.~Thouw,
``CORSIKA: A Monte Carlo Code to Simulate Extensive Air Showers,''
Forschungszentrum Karlsruhe Report FZKA 6019, 1998. [Online]. Available: 
http://www-ik.fzk.de/\verb#~#heck/corsika/

\bibitem{EGS}
W.~R.~Nelson, H.~Hirayama, and D.~W.~O.~Rogers, ``The EGS4 Code System,"
SLAC-0265, Dec. 1985.

\bibitem{GHEISHA}
H.~Fesefeldt, ``GHEISHA, the simulation of hadronic showers: physics and
applications,'' RWTH Aachen Report PITHA-85/02, 1985.

\bibitem{MOCCA} 
A.M. Hillas, ``Ground-based gamma-ray astronomy,''  Nuovo Cim. C16 pp.
701-712, 1996.

\bibitem{Scalzo2}
R.A. Scalzo, ``Observations of the EGRET Blazar W Comae with the Solar
Tower Atmospheric Cherenkov Effect Experiment,'' Ph.D. Thesis, Univ. of
Chicago, Chicago, Illinois, 2004.

\end{thebibliography}
\end{document}